\documentclass[twocolumn,letterpaper,prb,showpacs,amsmath]{revtex4}

\usepackage{graphicx,bm}

\newcommand{\im}{\mathrm i}

\begin{document}

\title{Selfconsistent gauge-invariant theory of in-plane infrared response of
high-T${}_\mathbf{c}$ cuprate superconductors involving spin fluctuations}

\author{Ji\v{r}\'{\i} Chaloupka}
\email[Electronic address:]{chaloupka@physics.muni.cz}
\affiliation{Department of Condensed Matter Physics, Faculty of Science,
Masaryk University, Kotl\'a\v{r}sk\'a 2, 61137 Brno, Czech Republic}

\author{Dominik Munzar}
\affiliation{Department of Condensed Matter Physics, Faculty of Science,
Masaryk University, Kotl\'a\v{r}sk\'a 2, 61137 Brno, Czech Republic}

\begin{abstract}
We report on results of our theoretical study of the in-plane infrared
conductivity  of the high-$T_c$ cuprate superconductors using the model where
charged planar quasiparticles are coupled to spin fluctuations.  The
computations include both the renormalization of the quasiparticles and the
corresponding modification of the current-current vertex function (vertex
correction), which ensures gauge invariance of the theory and local charge
conservation in the system.  The incorporation of the vertex corrections leads
to an increase of the total intraband optical spectral weight (SW) at finite
frequencies, a SW transfer from far infrared to mid infrared,  a significant
reduction of the SW of the superconducting condensate,  and an amplification
of characteristic features in the superconducting state spectra of the inverse
scattering rate $1/\tau$.  We also discuss the role of selfconsistency and
propose a new interpretation of a kink occurring in the experimental low
temperature spectra of $1/\tau$ around $1000\:\mathrm{cm}^{-1}$. 
\end{abstract}

\date{\today}


\maketitle

\section{INTRODUCTION}

Most of the existing calculations of the frequency dependent conductivity
$\sigma$ in high-$T_c$ cuprate superconductors (for representative examples
see Refs.~\onlinecite{
Quinlan:1996:PRB,Schachinger:1997:PRB,Munzar:1999:PhysicaC,
Carbotte:1999:Nature,Abanov:2001:PRB,Casek:2005:PRB,Hwang:2007:PRB}) employ
the approximation, where interactions of the excited states are not taken into
account, i.e., the so called vertex corrections (VC) in the perturbation
expansion for $\sigma$ are neglected.  It is well known that this approach is
not gauge invariant  and the results may depend  on the gauge of the vector
and scalar potentials. 
Physically, gauge invariant response is a manifestation of local charge
conservation in the system.
We recall that the usual restricted conductivity sum rule for the normal state
(NS)
\begin{equation}\label{NSCondSR}
I_f=-({\pi e^2}/{2\hbar^2})K\;,
\end{equation}
where 
\begin{equation}\label{FiniteomegaSW}
I_f=\int_{0^+}^\infty\sigma_{1x}(\omega)\mathrm{d}\omega
\end{equation}
is the intraband optical spectral weight and  
\begin{equation}\label{KE1}
K=-\frac{1}{V} \sum_{\bm{k} \alpha}
\frac{\partial^2 \varepsilon_{\bm{k}} }{\partial k_{x}^{2}} 
n_{\bm{k} \alpha}
\end{equation}
the so called effective kinetic energy per unit cell ($\alpha$ stands for the
spin index, $\varepsilon_{\bm{k}}$ is the dispersion relation, and $n_{\bm{k}
\alpha}$ the occupation factor), is a consequence of the gauge invariance
\cite{Nozieres:1999:book}.  For concreteness we have considered the
conductivity along the $x$-axis. The NS conductivity calculated using a no-VC
approach need thus not satisfy the sum rule. Similarly, for the
superconducting state (SCS), the spectral weight of the condensate $I_c$,
which is given by  
\begin{equation}\label{SCSCondSR}
I_f+I_c=-({\pi e^2}/{2\hbar^2})K\;,
\end{equation}
as obtained within a no-VC calculation, may be rather different from that of
the corresponding gauge-invariant approach. These possible problems call for
an assessment of the contribution of the relevant VC to $I_f$. 

Another reason for studying the VC has been highlighted by Millis and
coworkers.  If the VC were negligible, the infrared (IR) conductivity would be
determined solely by the quasiparticle selfenergy,  that can be, at least in
principle, extracted from photoemission data.  Millis and Drew
\cite{Millis:2003:PRB} found out, however,  that the effective scattering rate
in optimally doped Bi$_{2}$Sr$_{2}$CaCu$_{2}$O$_{8+\delta}$ obtained
using the selfenergy estimated from the photoemission data,  is much higher
than the value resulting from the IR data.  Similarly, Millis {\it et al.}
\cite{Millis:2005:PRB} demonstrated that the local selfenergy obtained by
fitting (without VC)  the IR data of thin films of
Pr$_{2-x}$Ce$_{x}$CuO$_{4+\delta}$ leads to much lower (by a factor of 3-4)
values of the quasiparticle velocity than observed by photoemission.  These
discrepancies appear to imply the presence of large VC. 

A brief account of the VC to the NS conductivity of cuprate superconductors
has been given by Monthoux and Pines in one of the pioneering papers on
the nearly antiferromagnetic Fermi liquid (NAFL) model
\cite{Monthoux:1994:PRB}. For this particular model, the VC have been found
to cause a slight increase (about 20\%) of resistivity.  Current vertex
renormalization  within the conserving fluctuation exchange (FLEX) approach
\cite{Manske} has been systematically investigated by Kontani and coworkers
(see, e.g., Refs.~\onlinecite{Kontani:2006:JPSJ} and
\onlinecite{Kontani:2005:condmat} and references therein).  They successfully
explained the anomalous temperature dependence of the Hall conductivity in
cuprates and estimated the contribution of the VC to the normal state
conductivity.  Benfatto {\it et al.}\cite{Benfatto:2005:PRB} and Aristov and
Zeyher \cite{Aristov:2005:PRB} discuss the role of the VC in the context of
optical response of systems possessing the $d$-density wave ground state, that
might occur in the pseudogap regime of underdoped cuprates.  We are not aware,
however, of any work incorporating the VC in computations of the optical
response in the superconducting state. 

The purpose of the present study is to explore the role played by the VC, both
in the NS and in the SCS state,  within the spin-fermion model, where charged
planar quasiparticles are coupled to spin fluctuations (SF).  We discuss the
violation, in the absence of VC, of the NS sum rule, the impact of the VC on
the dc conductivity, on the spectral weight of the superconducting condensate,
and on the shape of $\sigma_{1}(\omega)$.  An
important finding is that the incorporation of the VC leads to a
spectral-weight shift from far-infrared (FIR) to mid-infrared (MIR).  Finally, 
we address the related changes of sharp structures in the inverse scattering
rate $[1/\tau](\omega)$. 

The rest of the paper is organized as follows.  In Sec.~II we present the
basic equations of the theory,  the values of the input parameters,  and some
computational details.  Since part of the formalism  has been already detailed
in the related previous work by C{\'a}sek {\it et al.}\cite{Casek:2005:PRB} we
keep the account short.  Section \ref{secRD} contains our results and
discussion.  The quasiparticle Green's functions have been obtained here using
fully selfconsistent Eliashberg theory, whereas an approximate
non-selfconsistent approach, where the BCS propagator is corrected by the
coupling to the SF, has been adopted in Ref.~\onlinecite{Casek:2005:PRB}.  In
order to clarify the role of selfconsistency, we first neglect the VC and
compare the results of the two approaches (Sec.~\ref{secSF}).  In
Sec.~\ref{secVC} we then focus on the contribution of the vertex corrections.
The  summary and
conclusions are given in Sec.~\ref{secSM}. 

\section{THEORY, INPUT PARAMETERS, AND COMPUTATIONAL DETAILS}
\label{secTH}

Within the framework of the spin-fermion model, the electronic quasiparticles
are renormalized via a coupling to SF.  These are described by the spin
susceptibility $\chi_\mathrm{SF}(\bm{k},E)$ 
of the form motivated by neutron data 
and the electronic selfenergy
($2\times 2$ matrix) of the most sophisticated version of the theory is given
by \cite{Monthoux:1993:PRB}
\begin{equation}\label{selfE}
\Sigma(\bm{k},\im E)=\frac{g^2}{\beta N}\sum_{\bm{k'}, \im E'}
\chi_\mathrm{SF}(\bm{k}-\bm{k'},\im E-\im E')\,\mathcal{G}(\bm{k'},\im E') \;.
\end{equation}
Here $g$ is the coupling constant,  $\chi_\mathrm{SF}$ the Matsubara
counterpart of the spin susceptibility, and $\mathcal{G}$ is the Nambu
propagator of the renormalized electronic quasiparticles,  
$\mathcal{G}=\left[\im E\tau_0-(\epsilon_{\bm{k}}-\mu)\tau_3
-\Sigma(\bm{k},\im E)\right]^{-1}$.   
The sum runs through the Bloch vectors $\bm{k'}$ in the first Brillouin zone
($N$ in total) and the fermionic Matsubara energies $\im
E'=\im\pi(2n+1)/\beta$.  Note that equation \eqref{selfE} is a compressed form
of the Eliashberg equations. 

A popular approximation to the fully selfconsistent treatment of
Eq.~\eqref{selfE}, used, e.g., in
Refs.~\onlinecite{Quinlan:1996:PRB,Eschrig:2003:PRB,Casek:2005:PRB} consists in
starting with the BCS Nambu propagator with an estimated superconducting gap of
$d_{x^{2}-y^{2}}$ symmetry and including the lowest order correction due to the
coupling to the SF.  If we write the right hand side of Eq.~\eqref{selfE} as a
convolution, $\Sigma=g^2\chi_\mathrm{SF}\star\mathcal{G}$, then the latter
approach, called the hybrid approach in the following, yields  
$\Sigma_h=\Delta_{\bm{k}}\tau_{1}+g^2\chi_\mathrm{SF}\star\mathcal{G}_0$, 
where $\mathcal{G}_0$ stands for the BCS propagator with the gap
$\Delta_{\bm{k}}$.  This description is particularly suitable for situations,
where the coupling to the SF is not the only cause of superconductivity and
provides an \textit{additional} renormalization of the quasiparticles.  The
magnitude of the superconducting gap can be tuned by changing the input
amplitude $\Delta_{0}$ of $\Delta_{\bm{k}}$.  An important advantage of the
hybrid approach is that it avoids the necessity of numerical continuation
from Matsubara frequencies to the real axis.

Within the framework of the linear response theory, the optical conductivity
is given by the Kubo formula
\begin{equation}\label{sigma}
\sigma_{ij}(\bm{q}, \omega)=
\frac{(e^2/\hbar^2)K_{ij}+\Pi_{ij}(\bm{q}, \omega)}
     {\im(\omega+\im\delta)} \;,
\end{equation}
where $K_{ij}$ is the diamagnetic tensor, 
\begin{equation}\label{KE2}
K_{ij}=-\frac{1}{V} \sum_{\bm{k} \alpha}
\frac{\partial^2 \varepsilon_{\bm{k}} }{\partial k_i\partial k_j} 
n_{\bm{k} \alpha}\;. 
\end{equation}
In case of a simple cubic or square lattice 
and the nearest-neighbor tight-binding
dispersion relation, ${\rm Tr}K$ is proportional to the band energy (kinetic
energy), $\sum \varepsilon_{\bm{k}}n_{\bm{k}\sigma}$, per unit cell. For
this reason, $K$ is also called the effective kinetic energy. Further,
$\Pi_{ij}(\bm{q},\omega)$ in Eq.~\eqref{sigma} is the retarded correlation
function of the paramagnetic-current-density operator
\begin{equation}\label{jjcorr}
\Pi_{ij}(\bm{q},\omega)=\frac{\im}{\hbar}
\int_{-\infty}^{\infty}\mathrm{d} t\, \mathrm{e}^{\im\omega t}
\langle \left[j_i(\bm{q},t),j_j(-\bm{q},0)\right] \rangle 
\theta(t) \;.
\end{equation}

In the lowest order approximation, the correlator corresponds to a simple
bubble diagram  containing two independent quasiparticle lines.  In the limit
of small $\bm{q}$, its Matsubara counterpart can be written as
\begin{multline}\label{PiNV}
\Pi_{ij}(\bm{q},\im\hbar\nu)=-\frac{e^2}{\hbar^2}\frac{1}{V\beta}
\sum_{\bm{k},\im E} 
 \frac{\partial\varepsilon}{\partial k_i}
 \frac{\partial\varepsilon}{\partial k_j} \\
 \times \mathrm{Tr}\left[
  \mathcal{G}(\bm{k}+\bm{q}, \im E+\im\hbar\nu)
  \mathcal{G}(\bm{k},\im E)
 \right] \;.
\end{multline}
This is the approximation mentioned in the introduction, 
that is not gauge invariant and may lead to a violation 
of the sum rule \eqref{NSCondSR}. 

A general field theoretical method to overcome the problem of gauge invariance
was constructed by Nambu\cite{Nambu:1960:PR} in the context of the BCS theory.
In the present context it leads to the replacement of formula \eqref{PiNV}
with  
\begin{multline}\label{PiVC}
\Pi_{ij}(\bm{q},\im \hbar\nu)=-\frac{e^2}{\hbar^2}\frac{1}{V\beta} 
\sum_{\bm{k}, \im E}
\frac{\partial\varepsilon}{\partial k_i} \\
\times\mathrm{Tr}\,\left[
\mathcal{G}(\bm{k}+\bm{q},\im E+\im\hbar\nu)
\Gamma_j(\bm{k},\im E,\bm{q},\im\hbar\nu)
\mathcal{G}(\bm{k},\im E) \right] \;. 
\end{multline}
The bare vertex 
$\Gamma_j=(\partial\varepsilon/\partial k_j) \tau_0$ 
has been replaced with a renormalized one,  
$\Gamma_j(\bm{k},\im E,\bm{q},\im\hbar\nu)$ ($2\times 2$ matrix). 
For the theory to be gauge invariant, the renormalization of the vertex
representing the coupling of the quasiparticles to the electromagnetic field,
has to be consistent with the renormalization of the quasiparticles by the
interaction. It is known that $\Gamma$ has to obey the generalized Ward
identity, which is in fact a reformulated charge conservation
law\cite{Schrieffer:1988:Book}. The minimal vertex satisfying the generalized
Ward identity in the present case of the selfenergy given by Eq.~\eqref{selfE}
is the solution to the Bethe--Salpeter equation 
\begin{multline}\label{BSE}
\Gamma_j(\bm{k},\im E,\bm{q},\im\hbar\nu)=
\frac{\partial\varepsilon}{\partial k_j}\tau_0+
\frac{g^2}{N\beta}\sum_{\bm{k'},\im E'}
 \chi_{SF}(\bm{k}-\bm{k'},\im E-\im E') \\
 \times \mathcal{G}(\bm{k'}+\bm{q},\im E'+\im\hbar\nu)
 \Gamma_j(\bm{k'},\im E',\bm{q},\im\hbar\nu)
 \mathcal{G}(\bm{k'},\im E') \;.
\end{multline}
The second term on the right hand side of the equation is the VC.  The
correlator resulting from this gauge-invariant approach corresponds to the sum
of all ladder diagrams, where non-crossing SF lines connecting the two
quasiparticle lines are inserted into the conductivity bubble. 

Formally, the hybrid approach can be made gauge invariant as well by using a
somewhat simpler form of the renormalized vertex.  The results, however, are
qualitatively different from those of the selfconsistent approach.  

The input quantities of the theory are the dispersion relation
$\varepsilon_{\bm{k}}$, the chemical potential (or the electron density), the
spin susceptibility $\chi_{\mathrm{SF}}$, and the coupling constant $g$.  We
have used the second nearest neighbor tight-binding dispersion relation and
the model spin susceptibility of the same form as in
Ref.~\onlinecite{Casek:2005:PRB}, containing the resonance mode and a
continuum  with dimensionless spectral weights of $0.01b_{M}$ and $0.01b_{C}$,
respectively. The values of all input parameters are given in Table
\ref{tab:param}. They are the same as in Ref.~\onlinecite{Casek:2005:PRB},
except for $g$ (${1.52{\rm\,eV}}$ in
Ref.~\onlinecite{Casek:2005:PRB}).  The present value of $g$ of ${2{\rm\,eV}}$
yields $T_c=64{\rm\,K}$ and $\Delta=20\:\mathrm{meV}$ ($\Delta$ is the
amplitude of the gap).  In the computations of Sec.~III B, slightly different
values of $t$, $n$, and $g$ have been used, in Table \ref{tab:param}
they are given in the brackets.
\cite{Andersen:1995:JPCS}.
For $g=2\:\mathrm{eV}$ ($g=3\:\mathrm{eV}$) we obtain 
$T_c=77\:\mathrm{K}$ and $\Delta=22\:\mathrm{meV}$ 
($T_c=89\:\mathrm{K}$ and $\Delta=27\:\mathrm{meV}$).

\begin{table*}
\begin{ruledtabular}
\begin{tabular}{cccccccccccccc}
$a$ [\AA] & 
$d$ [\AA] & 
$t [\mathrm{eV}]$ & 
$t' [\mathrm{eV}]$ & 
$n$ & 
$g [\mathrm{eV}]$ &
$\hbar\omega_0 [\mathrm{eV}]$ & 
$\Gamma [\mathrm{eV}]$ &
$\xi [a]$ &
$\hbar\omega_C [\mathrm{eV}]$ &
$\Gamma_C [\mathrm{eV}]$ &
$\xi_C [a]$ & 
$b_M$ & $b_C$ \\
\hline
3.828 & 11.650 & 0.250 (0.350) & -0.100 & 0.76(0.82) & 2.0(2.0/3.0) &
0.040 & 0.010 & 2.35 & 0.400 & 1.000 & 0.52 & 1 & 4
\end{tabular}
\end{ruledtabular}
\caption{Values of the parameters used in the computations. 
Lattice parameters are denoted by $a$ and $d$,   
$t$ and $t'$ are the parameters of the dispersion relation, 
$n$ is the number of electrons per unit cell, 
the meaning of the parameters $\omega_{0}$, $\Gamma$, $\xi$, 
$\omega_{C}$, $\Gamma_{C}$, and $\xi_{C}$ specifying the spin susceptibility 
is explained in Ref.~\onlinecite{Casek:2005:PRB}. 
For the normal state, we take  
$\Gamma=0.070\:\mathrm{eV}$ and
$\xi=1.57a$.
The value of the input amplitude $\Delta_{0}$ of $\Delta_{{\bm k}}$
used within the hybrid-approach was adjusted to yield the same 
amplitude of the resulting gap $\Delta$ as the selfconsistent approach. 
In the computations of Sec.~III B, 
the values in the brackets have been used, 
the spectra shown in the figures 
\ref{fig:spect}, \ref{fig:corr}, \ref{fig:gdep}, \ref{fig:Wfun}
have been obtained with $g=3\:\mathrm{eV}$.} 
\label{tab:param} 
\end{table*}

Note, that the present formulation assumes a single band only.  In order to
compare our results with experimental data on bilayer cuprates, we multiply
the conductivity and related quantities (like $K$) by a factor $N_\mathrm{pl}=2$
reflecting the two planes in the unit cell. To address this issue carefully,
we have also performed calculations with a model containing two bands --
bonding and antibonding. The spin susceptibility was redistributed between
the odd- and even-channels, the resonant mode being active in the odd channel
\cite{Fong:2000:PRB}. The single-particle part of our computations is similar
to that of Ref.~\onlinecite{Eschrig:2002:PRL}. We have neglected the VC for
simplicity.  The results indicate, that for reasonable values of the
intrabilayer hopping matrix elements (of the order of $100\:\mathrm{meV}$), the
band splitting has a negligible effect on the in-plane conductivity, changing
only the relative contributions of the bonding and the antibonding band to the
resulting spectra.

The iterative solution of the selfenergy equation \eqref{selfE} and
especially of the Bethe-Salpeter equation \eqref{BSE} involves many
convolutions both in Matsubara frequencies and in $\bm{k}$-space.  Together
with the requirement of high accuracy data for the subsequent analytical
continuation to the real axis, this leads to a computationally very demanding
task.  We have performed the convolutions using the FFT algorithm taking the
full advantage of the symmetries of $\Sigma$ and $\Gamma$.  Typically, we have
used a grid of $96\times96$ points in the Brillouin zone and a cutoff of
$10\:\mathrm{eV}$ (approximately $3.5$ times the bandwidth) to limit the
number of Matsubara frequencies.  We have checked, by varying the density of
the grid and the cutoff, that these values are sufficient.  The analytical
continuations to the real axis were performed using the standard method of
Pad\'{e} approximants\cite{Vidberg:1977:JLTP}.

\section{RESULTS AND DISCUSSION}
\label{secRD}

\subsection{Role of selfconsistency}
\label{secSF}

\begin{figure}[tbp]
\includegraphics[width=8.2cm]{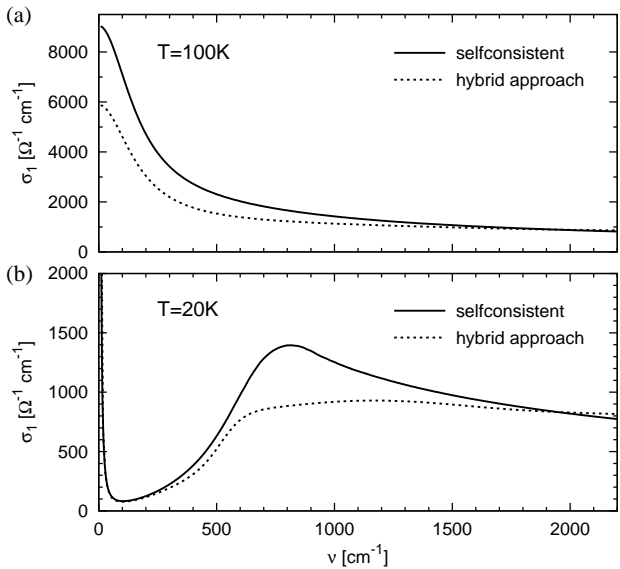}
\caption{
Real part of the optical conductivity as obtained by using the selfconsistent
approach (solid lines) and the hybrid approach (dashed lines) for the normal
state at $T=100\:\mathrm{K}$ and for the superconducting state at
$T=20\:\mathrm{K}$. 
}
\label{fig:hybsig}
\end{figure}

Here we compare the conductivity spectra computed using the
(nonselfconsistent) hybrid approach with those of the selfconsistent
Eliashberg theory.  Figure~\ref{fig:hybsig} shows the spectra of $\sigma_{1}$
for the NS at $T=100\:\mathrm{K}$ (a) and for the SCS at $T=20\:\mathrm{K}$
(b).  For the NS, the conductivity profiles are quite featureless. The
dc value of the hybrid approach is by about 30\% smaller.  This is because the
hybrid approach, which is -- only for the NS -- equivalent to the first
iteration of Eq.~\eqref{selfE}, overestimates the magnitude of the
quasiparticle selfenergy.

\begin{figure}[tbp]
\includegraphics[width=8.2cm]{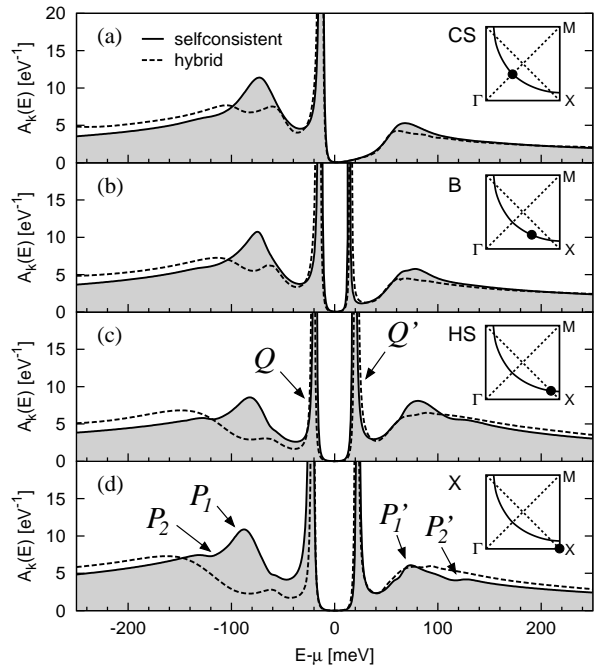}
\caption{
Quasiparticle spectral functions for selected $\bm{k}$-points defined in
Ref.~\onlinecite{Casek:2005:PRB} and shown in the insets as obtained by using
the selfconsistent approach (solid lines) and the hybrid approach (dashed
lines)  for the superconducting state at $T=20\:\mathrm{K}$.  Our notation of
spectral structures is introduced in panels (c) and (d). 
}
\label{fig:hybAkE}
\end{figure}

Next we discuss the spectra for the SCS. They display an increase of
$\sigma_{1}$ starting at low frequencies, which becomes steeper around
$\hbar\omega_{0}+\Delta$ as discussed in Ref.~\onlinecite{Casek:2005:PRB}.
The character of the maximum following the onset, however, depends on the
level of theory used. For the hybrid approach, the conductivity exhibits a
very broad maximum, whereas the selfconsistent approach yields a relatively
sharp maximum near $800\:\mathrm{cm}^{-1}$, consistent with experimental data
\cite{Boris:2004:Science,Marel:2003:Nature,Timusk:2004:Nature}. The origin of
this difference can be traced back to that of the quasiparticle spectral
functions, using the method presented in Ref.~\onlinecite{Casek:2005:PRB}.  It
is based on the formula valid for the NS ($\bm{q}=0$)
\begin{multline}\label{convAkE}
\sigma_1(\omega)\propto \frac1\omega\sum_{\bm{k}}
\frac1\hbar\frac{\partial\varepsilon}{\partial k_i}
\frac1\hbar\frac{\partial\varepsilon}{\partial k_j} \\
\int A(\bm{k},E)A(\bm{k},E+\hbar\omega)[n_F(E)-n_F(E+\hbar\omega)]\mathrm{d}E\;,
\end{multline}
where $A(\bm{k},E)$ is the spectral function and $n_F$ the Fermi function,
and a similar formula involving the matrix spectral function valid for the
SCS. These formulas allow one to understand the structures in the spectra
of $\sigma_1$ in terms of ``transitions'' between various components of $A$.
It has been shown in Refs.~\onlinecite{Munzar:1999:PhysicaC,Casek:2005:PRB},
that the onset of $\sigma_1$ in the SCS including the maximum is determined by
transitions between the quasiparticle peaks and the incoherent parts of $A$.
The spectral functions for selected $\bm{k}$-points are presented in
Fig.~\ref{fig:hybAkE}. 
A detailed discussion of the spectral structures obtained at the hybrid level
can be found in Ref.~\onlinecite{Casek:2005:PRB}.  Here we concentrate on the
comparison  between the results of the selfconsistent and the hybrid approach.
The positions of the quasiparticle peaks $Q$, $Q'$ are identical,  which has
been achieved by tuning the input BSC gap of the hybrid approach.  The maxima
$P_{1}$ and $P'_{1}$ of the incoherent part in the selfconsistent case are
located closer to the quasiparticle peaks and are considerably sharper.  This
causes the steeper onset and the sharper maximum of $\sigma_1$.  It can be
seen that this maximum  simply reflects 
the maxima of the incoherent part of $A$ -
$P_{1}$ and $P'_{1}$. 

\subsection{Vertex corrections}
\label{secVC}

In this section we focus on the central topic of the paper of how the model
spectra change when the VC are included. 
\begin{figure}
\includegraphics[width=8.2cm]{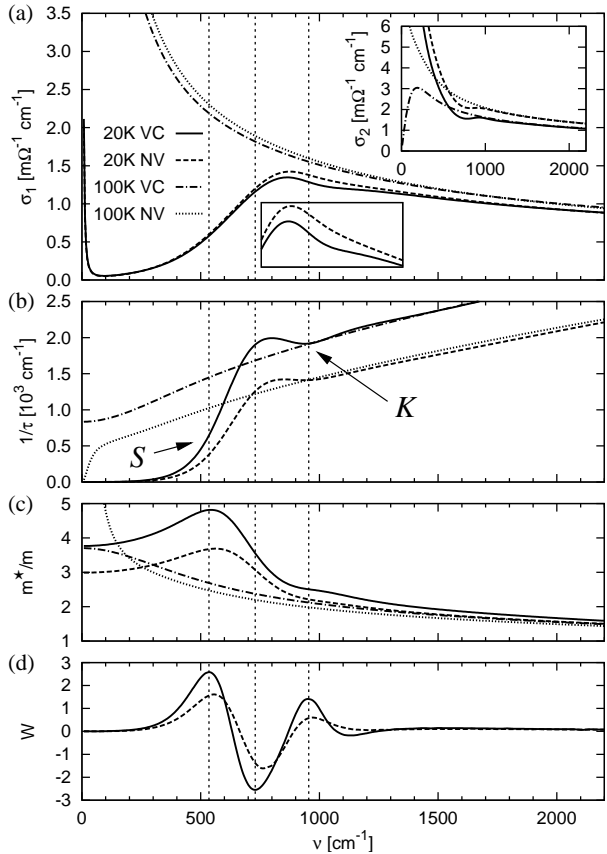}
\caption{
The spectra of the optical conductivity, the inverse scattering rate, the mass
enhancement factor, and the function $W$ calculated with the vertex
corrections neglected (NV) and with the vertex corrections included (VC) for
the set of input parameters leading to realistic values of $T_c$ and $\Delta$.
The insets of panel (a) shows the vertically zoomed area near the conductivity
maximum and the spectra of $\sigma_{2}$. In panel (d), only the
superconducting state spectra are shown. The vertical dashed lines are drawn
to guide the eye.  A notation of spectral structures is introduced in (b). 
}
\label{fig:spect}
\end{figure}
Figure~\ref{fig:spect} shows the optical conductivity and three
related quantities: the inverse scattering rate $[1/\tau](\omega)$ and the
mass enhancement factor $[m^*/m](\omega)$, that are defined by the
extended-Drude-model formula 
\begin{equation}
\sigma(\omega)=\frac{\epsilon_0\omega_\mathrm{pl}^2}
{[1/\tau](\omega)-\im\omega[m^*/m](\omega)} \;, 
\end{equation}
where $\omega_{\rm pl}$ is the plasma frequency, and 
$W(\omega)=(2\pi)^{-1}\mathrm{d}^2[\omega/\tau(\omega)]/\mathrm{d}\omega^2$.
For the NS of a weakly coupled isotropic electron-phonon system 
the function $W(\omega)$ is approximately equal 
to the electron-phonon spectral density \cite{Marsiglio:1998:PhysLett,
Schulga:2001:book}.
The solid and the dashed lines correspond to the SCS at $T=20\:\mathrm{K}$,
the dashed-dotted and the dotted lines to the NS at $T=100\:\mathrm{K}$. The
solid and the dashed-dotted lines (the dashed and the dotted lines) correspond
to the selfconsistent computation with (without) the VC. The values of the
dc conductivity $\sigma_\mathrm{dc}$, the effective kinetic energy $K$
(multiplied by $d$ to obtain the dimension of energy), the spectral weight at
finite frequencies $I_f$, and that of the condensate, $I_c$, are given in
Table~\ref{tab:weight}. The values of $I_f$ and $I_c$ have been multiplied by
$2\hbar^2 d/\pi e^2$ to allow a direct comparison with those of $Kd$.

\begin{table}
\begin{ruledtabular}
\begin{tabular}{c|cc|ccc}
\multicolumn{6}{l}{
 $g=2\:\mathrm{eV}$ ~~~~~~
 $T_c=77\:\mathrm{K}$ ~
 $\Delta_X=22\:\mathrm{meV}$ ($20\:\mathrm{K}$)}\\
\hline
 & \multicolumn{2}{l|}{$T=20\:\mathrm{K}$} &
   \multicolumn{3}{l}{$T=100\:\mathrm{K}$} \\
 & \multicolumn{2}{l|}{$Kd=-0.44688\:\mathrm{eV}$} &
   \multicolumn{3}{l}{$Kd=-0.44653\:\mathrm{eV}$} \\
 & $I_f$ [eV] & $I_c$ [eV]  & $I_f$ [eV]  & 
 $I_c/|K|$ [\%] & $\sigma_\mathrm{dc} [\Omega^{-1}\mathrm{cm}^{-1}]$ \\
\hline
NV     & 0.27079 & 0.17609 & 0.43205 & 3.2 & $1.62\cdot 10^4$ \\
VC     & 0.29752 & 0.14936 & 0.44624 & 0.1 & $1.41\cdot 10^4$ \\
\hline
\hline
\multicolumn{6}{l}{
 $g=3\:\mathrm{eV}$ ~~~~~~
 $T_c=89\:\mathrm{K}$ ~
 $\Delta_X=27\:\mathrm{meV}$ ($20\:\mathrm{K}$)}\\
\hline
 & \multicolumn{2}{l|}{$T=20\:\mathrm{K}$} &
   \multicolumn{3}{l}{$T=100\:\mathrm{K}$} \\
 & \multicolumn{2}{l|}{$Kd=-0.40741\:\mathrm{eV}$} &
   \multicolumn{3}{l}{$Kd= -0.40699\:\mathrm{eV}$} \\
 & $I_f$ [eV] & $I_c$ [eV]  & $I_f$ [eV]  & 
 $I_c/|K|$ [\%] & $\sigma_\mathrm{dc} [\Omega^{-1}\mathrm{cm}^{-1}]$ \\
\hline
NV     & 0.28871 & 0.11870 & 0.38580 & 5.2 & $8.0\cdot 10^3$ \\
VC     & 0.31312 & 0.09429 & 0.40699 & 0.0 & $7.5\cdot 10^3$ \\
\end{tabular}
\end{ruledtabular}
\caption{\label{tab:weight}
Values of the effective kinetic energy $K$, the spectral weight at finite
frequencies $I_f$, and the spectral weight of the singular component
(condensate) $I_c$ computed without vertex corrections (NV) and with the
vertex corrections included (VC). The data are presented for two values of
the coupling constant $g$, for the superconducting state at $T=20\:\mathrm{K}$
and for the normal state at $T=100\:\mathrm{K}$.  For the normal state, the
values of the dc conductivity are also given.
}
\end{table}

We begin our discussion with the NS.  It can be seen in Table~\ref{tab:weight}
that in the absence of VC the sum rule $I_f\sim K$ \eqref{NSCondSR} is not
fulfilled.  It means that the conductivity possesses an unphysical singular
component with the spectral weight $I_c$ determined by Eq.~\eqref{SCSCondSR}.
The values of $I_c$ of ca $3\%$ (ca $5\%$) of $-K$ for $g=2\:\mathrm{eV}$
($g=3\:\mathrm{eV}$) are small but significant.  The unphysical component
manifests itself also in the spectra of related quantities - see the spectra
of $\sigma_{2}$ in the inset of Fig.~\ref{fig:spect} (a), the drop of $1/\tau$
at low frequencies [Fig.~\ref{fig:spect} (b)], and the corresponding
divergence of $m^{*}/m$ [Fig.~\ref{fig:spect} (c)].  With the VC included the
sum rule is satisfied.  The spectral weight increase due to the VC [$\Delta
I_f(\mathrm{VC})$] is equal to $I_c$ (within the numerical error related
to discrete $\bm{k}$-sampling).  Based on the relatively small values of
$I_c$, the changes of the spectra due to the VC can be expected to be small.
Indeed, only a slight decrease of $\sigma_{1}$ in the FIR can be observed in
Fig.~\ref{fig:spect}.  The dc conductivity also decreases (see 
Table \ref{tab:weight}) which is consistent with the results of Monthoux and
Pines \cite{Monthoux:1994:PRB}.  This trend, however, is not universal, as
will be discussed in Sec.~\ref{sec:transf}. 

In the SCS, the VC increase $I_f$, which leads to a reduction of $I_c$ (see
Table~\ref{tab:weight}).  This effect is explored in detail in
Sec.~\ref{sec:IcVC}.  The real part of the conductivity is affected mainly
near then maximum around $800\:\mathrm{cm}^{-1}$: the VC make it slightly
sharper.  The changes are further amplified in the spectra of related
quantities $1/\tau$ and $W$ to be discussed in Sec.~\ref{sec:finedet}.

\subsubsection{Spectral weight transfer from FIR to MIR}
\label{sec:transf}

\begin{figure}
\includegraphics[width=8.2cm]{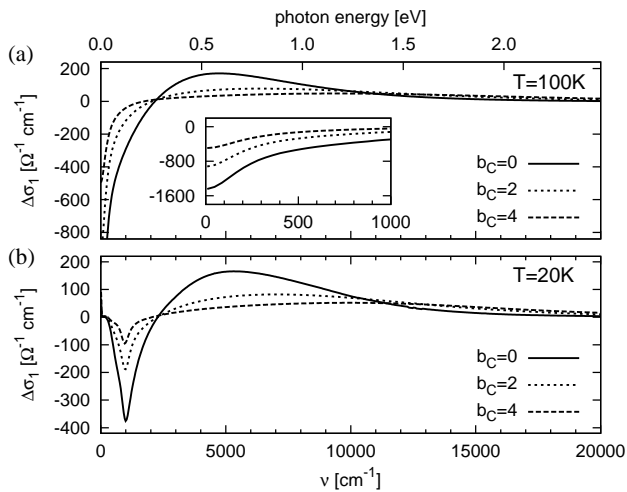}
\caption{
(a) Contribution of the VC to the real part of the normal state conductivity
at $T=100\:\mathrm{K}$ for various values of the relative spin-continuum
weight $b_C$. The low energy parts of the spectra are shown in the inset.
(b) The same for the superconducting state at $T=20\:\mathrm{K}$. 
}
\label{fig:corr}
\end{figure}

Figure~\ref{fig:spect}(a) shows a decrease of $\sigma_1$ in FIR with the
incorporation of the VC which seems to be inconsistent with the increase of
the finite-frequency spectral weight $I_f$.  A resolution of this apparent
controversy is provided by Fig.~\ref{fig:corr}, which shows the contribution
$\Delta\sigma_{1}(\mathrm{VC})$ of the VC to $\sigma_{1}$ in a wide spectral
range.  It can be seen that $\Delta\sigma_{1}(\mathrm{VC})$ is negative for
$\omega<2000\:\mathrm{cm}^{-1}$ but positive  for
$\omega>2000\:\mathrm{cm}^{-1}$.  The magnitude of the contribution of the
latter spectral range to $I_f$ is larger than that of the former, which leads
to the total increase of $I_f$, both for the NS and for the SCS.  For the SCS,
the magnitude of $\Delta\sigma_{1}(\mathrm{VC})$ becomes small as the
frequency approaches zero.  This is due to the vanishing real part of 
conductivity in this limit.  Figure~\ref{fig:corr} further demonstrates that
the maximum of $\Delta\sigma_{1}(\mathrm{VC})$ shifts towards higher
frequencies and becomes broader with increasing weight 
of the spin fluctuation continuum. 

In order to explain the behavior displayed in Fig.~\ref{fig:corr}, we use an
extension of the formula \eqref{convAkE} that applies to the theory involving
the VC.  In the generalized formula, which has been obtained by manipulations
starting from Eq.~\eqref{PiVC}, the integral on the right hand side of
Eq.~\eqref{convAkE} is replaced with a more general convolution of the
spectral functions of the form
$\int\mathrm{d}E'\int\mathrm{d}E'' A(\bm{k},E')A(\bm{k},E'')
\Xi(E',E'',\hbar\omega)$,
with the the Matsubara counterpart of the convolution kernel given by
\begin{equation}
\Xi(E',E'',\im\hbar\nu)=\frac1{\hbar\beta}\,\mathrm{Im}\sum_{\im E} 
\frac{\Gamma(\bm{k},\im E,\im\hbar\nu)}
{(\im E+\im\hbar\nu-E')(\im E-E'')} \;.
\end{equation}
If we use the bare vertex $\Gamma=\partial\varepsilon/\partial k$, 
we arrive at the relation 
$\Xi\propto\delta(E''+\hbar\omega-E')[n_F(E'')-n_F(E''+\hbar\omega)]$ 
which, after performing one of the integrations in the convolution, provides
exactly the formula \eqref{convAkE}.  In contrast, any kernel $\Xi$
corresponding to a frequency dependent renormalized vertex will be broader
than the delta function and will result in a broadening of the conductivity
profile. 

The effect of the VC on the conductivity can be vaguely viewed as consisting
of two ingredients: (a) the increase of the overall spectral weight $I_f$ and
(b) the broadening of the conductivity profile, which causes a transfer of
spectral weight towards higher frequencies.  The point (b) accounts for the
trends shown in Fig.~\ref{fig:corr}.  The change of the dc conductivity is
determined by a competition of (a) and (b).  For our choice of the input
parameters, the VC reduce the dc conductivity, in some cases, however, where
$\Delta I_f(\mathrm{VC})$ is large, an increase of $\sigma_\mathrm{dc}$ may
occur.  Some examples can be found in Ref.~\onlinecite{Kontani:2005:condmat}. 

The spectral weight redistribution of Fig.~\ref{fig:corr} is similar to but
considerably smaller than the one that could be expected based on the
arguments by Millis {\it et al}\cite{Millis:2005:PRB}.  Note, however, that
the discrepancy between the value of the quasiparticle velocity along the
Brillouin zone diagonal resulting from the present computations of
$1.1\:\mathrm{eV\AA}$ for $g=2\:\mathrm{eV}$ ($0.75\:\mathrm{eV\AA}$ for
$g=3\:\mathrm{eV}$) and the experimental value for YBCO of ca
$1.6\:\mathrm{eV\AA}$ \cite{Borisenko:2006:PRL} is not as dramatic as in
Ref.~\onlinecite{Millis:2005:PRB}. 

\subsubsection{Effect of the vertex corrections on the spectral 
weight of the condensate}
\label{sec:IcVC}

We have already noted that the VC significantly increase the values of $I_f$
and reduce those of $I_c$.
\begin{figure}[tp]
\includegraphics[width=8.2cm]{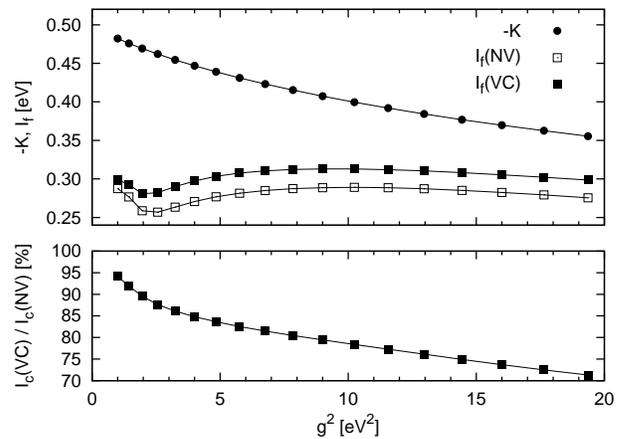}
\caption{
(a) The coupling-constant dependence of the effective kinetic energy and of
the intraband spectral weight obtained with the vertex corrections neglected
(NV) and with the vertex corrections included (VC).  (b) The coupling-constant
dependence of the ratio of the condensate weight obtained with the vertex
corrections included to that obtained using the vertex corrections neglected.
}
\label{fig:gdep}
\end{figure}
This is further documented in Fig.~\ref{fig:gdep}, which shows $I_f$ and $I_c$
as functions of the coupling constant $g$.  For values of $g^2$ of
$4-10\:\mathrm{eV}^{2}$, leading to reasonable values of $T_c$ and $\Delta$,
the magnitude of $\Delta I_f(\mathrm{VC})$ is about 6\% of that of $K$.  It
can be seen in part (b) that the corresponding change in the condensate weight
$I_c$ is up to 20\%.  At low values of $g$, $\Delta I_f(\mathrm{VC})$ scales
with $g^{2}$.  This can be easily interpreted by considering the diagram
series for the correlator \eqref{PiVC}.  The contribution of a diagram with
$N$ spin-fluctuation lines connecting the quasiparticle lines in the bubble is
proportional to $g^{2N}$.  In the limit of small $g$, only the lowest-order
diagrams ($N=0$ and $N=1$) survive, leading to the observed behavior.

The temperature dependence of the effective kinetic energy $K$ and the band
energy
$KE=\sum\epsilon_{\bm{k}} n_{\bm{k}\sigma}$ 
obtained within Eliashberg theory with SF, was already discussed in
Ref.~\onlinecite{Schachinger:2005:PRB}.  Our approach yields a similar
behavior of the two quantities.  The resulting value of 
$K(\mathrm{NS}, T=20\:\mathrm{K})-K(\mathrm{SCS}, T=20\:\mathrm{K})$
is also similar to that obtained by C{\'a}sek {\it et al.}
\cite{Casek:2005:PRB} using the hybrid approach.  The analysis published in
Ref.~\onlinecite{Casek:2005:PRB}, however, cannot be easily extended to the
present fully-selfconsistent theory.  The  contribution $\Delta
I_f(\mathrm{VC})$ of the VC to $I_f$ is only weakly temperature dependent
above $T_c$.  Below $T_c$, $\Delta I_f(\mathrm{VC})$ slightly decreases,
exhibits a minimum, and then increases.  The low temperature value is somewhat
higher than that of the NS. 

\subsubsection{Spectral structures of $[1/\tau](\omega)$ and $W(\omega)$
for the superconducting state}
\label{sec:finedet}

Having discussed the global trends of the spectral changes due to the VC, we
concentrate here on pronounced features in the spectra of $1/\tau$ and $W$.
The SCS spectra of $1/\tau$ shown in Fig.~\ref{fig:spect} exhibit the famous
onset starting around the frequency of the resonance mode $\omega_{0}$,
becoming steeper around $500\:\mathrm{cm}^{-1}$ (this feature is labeled as
$S$),  and reaching a sharp maximum around $800\:\mathrm{cm}^{-1}$.
Surprisingly, the maximum is followed by a kink labeled as $K$.  Note that the
spectra are fairly similar to the experimental ones 
of optimally doped materials 
\cite{Boris:2004:Science,Marel:2003:Nature,Timusk:2004:Nature}.  All these
features appear already in the spectra of $\sigma_{1}$, 
however, they are more pronounced in those of $1/\tau$.
The function $W$ is approximately
proportional to the second derivative of $1/\tau$ and can thus be expected to
possess two maxima corresponding to the structures $S$ and $K$ and a minimum
close to the maximum of $1/\tau$.  This is indeed the case, as shown in
Fig.~\ref{fig:spect} (d).

The origin of the feature $S$ has been elucidated by C\'{a}sek {\it et
al}\cite{Casek:2005:PRB}. It is due to the appearance above 
$\hbar\omega_{0}+\Delta$ of excitations of the nodal region, 
consisting of a nodal quasiparticle, an antinodal quasiparticle, 
and the resonance mode. 
The arguments of Ref.~\onlinecite{Casek:2005:PRB}, 
even though formulated at the level of the hybrid approach, 
remain to be valid also in the context of the
present fully selfconsistent theory.  As noted for the first time by Carbotte,
Schachinger, and Basov \cite{Carbotte:1999:Nature},  the energy of the
relevant maximum in $W$ is close to $\hbar\omega_{0}+\Delta$.

The sharp maximum of $1/\tau$ and the corresponding minimum of $W$ can be
shown to result from transitions $P_{1}\rightarrow Q'$ and $Q\rightarrow
P'_{1}$ (see Fig.~\ref{fig:hybAkE}).  The energy of the structure is close to
$\hbar\omega_{0}+2\Delta$, which can be understood using the arguments
presented in Ref.~\onlinecite{Casek:2005:PRB}.  The presence of this
characteristic energy scale in the optical spectra of the high-$T_c$ cuprates
has been for the first time predicted by Abanov, Chubukov, and Schmalian
\cite{Abanov:2001:PRB}.  In their work, the corresponding spectral structure
is attributed to processes involving a Bogoljubov quasiparticle with energy
$\Delta$ and a sharp onset of the incoherent part of $A(\bm{k},E)$ at
$\Delta+\hbar\omega_{0}$. 

The kink labeled as $K$ can also be interpreted in terms of the quasiparticle
spectral functions.  Note that the maxima $P_{1}$ and $P'_{1}$ in
Fig.~\ref{fig:hybAkE} that are connected to the maximum of $\sigma_{1}$ and
the related structures in the spectra of $1/\tau$ and $W$, are followed by
shoulder features, labeled as $P_{2}$ and $P'_{2}$,  on their high-energy
sides.  They can be expected, based on Eq.~\eqref{convAkE}, to manifest
themselves also in the conductivity.  Our detailed calculations show that (a)
the shoulder features are indeed responsible for the kink $K$ in $1/\tau$ and
(b) they can be attributed to excited states involving two magnetic
excitations.  We recall that the maxima $P_{1}$ and $P'_{1}$ correspond to
excited states involving one Bogoljubov quasiparticle and just one magnetic
excitation.  The characteristic energy of the shoulder features is
$2\hbar\omega_{0}+\Delta$ and that of the kink $K$ (which can be associated
with transitions $P_{2}\rightarrow Q'$ and $Q\rightarrow P'_{2}$) is
$2\hbar\omega_{0}+2\Delta$.  Abanov, Chubukov, and Schmalian
\cite{Abanov:2001:PRB} also predicted a structure located at
$2\hbar\omega_{0}+2\Delta$.  Their interpretation, however, differs from 
ours.  In their theory, the structure is due to transitions between
negative- and positive- energy satellites (incoherent components) of $A$. 

In order to check the proposed assignment of the spectral structures, 
we have studied the $\omega_{0}$-dependence of $1/\tau$ and $W$. 
\begin{figure}
\includegraphics[width=8.2cm]{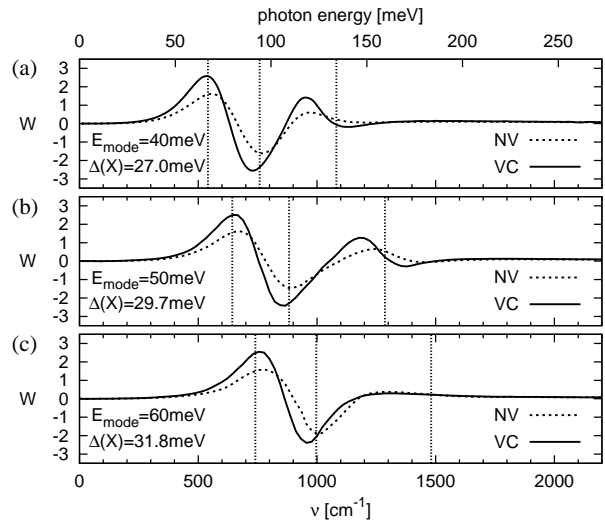}
\caption{
The function $W$ for the superconducting state at $T=20\:\mathrm{K}$ obtained
with the vertex corrections neglected (NV) and with the vertex corrections
included (VC) for three values of the energy of the resonant mode in the spin
susceptibility.  The vertical lines are located at the characteristic energies
of 
$\hbar\omega_0+\Delta$, $\hbar\omega_0+2\Delta$, and $2\hbar\omega_0+2\Delta$.  
}
\label{fig:Wfun}
\end{figure}
Figure~\ref{fig:Wfun} shows the spectra of $W$ for three values of the energy
of the magnetic mode.  It can be seen that the first maximum and the minimum
are located close to  $\hbar\omega_0+\Delta$ and $\hbar\omega_0+2\Delta$,
respectively, in agreement with the above considerations.  The second maximum
corresponding to the kink is, in the absence of VC, located somewhat below
$2\hbar\omega_0+2\Delta$.  

Finally, we address the role of the VC.  They lead (a) to an increase in the
amplitude of the structures of $1/\tau$ and $W$,  which is due to the combined
effect of the significant decrease of $\sigma_{2}$ [see the inset of
Fig.~\ref{fig:spect} (a)] and a sharpening of the features in $\sigma_{1}$, and
(b), to a slight shift of the structures towards lower energies.  The shift may
be partially caused by a weak attractive interaction between the
quasiparticles leading to an excitonic effect. 

\section{SUMMARY AND CONCLUSIONS}
\label{secSM}

The changes of the infrared conductivity caused by the vertex corrections (VC)
are not dramatic, which provides a justification for earlier computations,
where these corrections were neglected.  Some aspects of the changes, however,
appear to be important. \\
(i) The normal state conductivity computed without the VC does not satisfy the
restricted sum rule. The calculated spectral weight at finite frequencies
($I_f$) is by a few percent lower than the value dictated by the sum rule.
The incorporation of the VC leads to the increase of $I_f$, removing this
discrepancy. \\
(ii) The increase of $I_f$ is associated with a broadening of $\sigma_{1}$.
The far-infrared conductivity (including the dc value) may decrease or
increase depending on the magnitude of the change of $I_f$ and the degree of
the broadening; at high frequencies the conductivity increases.  For our
values of the input parameters, $\sigma_{1}$ decreases (increases) below
(above) ca $2000\:\mathrm{cm}^{-1}$. \\
(iii) The increase of $I_f$ occurs also for the superconducting state.  Since
the sum of $I_f$ and the spectral weight of the superconducting condensate
remains constant, this implies a reduction of the latter.  For relevant values
of the input parameters, the weight of the condensate decreases by 15-20\%. \\
(iv) The VC lead to an amplification of the characteristic features in the
superconducting state spectra of the inverse scattering rate $1/\tau$, that
have been used to support the spin-fluctuation scenario: the onset of $1/\tau$
around $500\:\mathrm{cm}^{-1}$ becomes steeper and the maximum around
$800\:\mathrm{cm}^{-1}$ more pronounced.  

In addition to studying the changes brought about by the VC, we have also
investigated the role of selfconsistency by comparing the results obtained
using the selfconsistent Eliashberg equations with those of the
nonselfconsistent hybrid approach (i.e., approximately, the first iteration of
the equations). The main results are (a) the hybrid approach considerably
overestimates the magnitude of the quasiparticle selfenergy, which leads to
lower values of the conductivity in far-infrared,  and (b) some spectral
features, in particular the sharp maximum in the superconducting state spectra
centered at ca $800\:\mathrm{cm}^{-1}$ and the kink at ca
$1000\:\mathrm{cm}^{-1}$   appear only at the selfconsistent level.  With the
aid of the quasiparticle spectral function $A$, we attribute the kink to the
onset of transitions involving incoherent satellites of $A$ corresponding to
states with doubly excited resonance mode.
\footnote{Note, that multiple excitations of the resonance mode are not
included at the hybrid level.}

The computed spectra are in reasonable agreement with experimental data of
optimally doped cuprates,  including such details as the shape of the maximum
in the inverse scattering rate.  In addition, the theory allows one to
interpret most of the features of the data in terms of Bogoljubov
quasiparticles and magnetic excitations.  In the superconducting state, a
crucial role is played by the resonance mode.  We are not aware of any
comparable interpretation in terms of electron-phonon coupling. 

\section*{ACKNOWLEDGMENTS}

This work was supported by the Ministry of Education of Czech Republic
(MSM0021622410). J.~Ch.~thanks B.~Keimer and G.~Khaliullin for their
hospitality during a stay at MPI Stuttgart. We gratefully acknowledge helpful
discussions with J.~Huml\'{\i}\v{c}ek, C.~Bernhard, A.V.~Boris, N.N.~Kovaleva
and B.~Keimer.


\begin{thebibliography}{99}

\bibitem{Quinlan:1996:PRB}
S.~M.~Quinlan, P.~J.~Hirschfeld, and D.~J.~Scalapino, 
Phys.~Rev.~B \textbf{53}, 8575 (1996). 

\bibitem{Schachinger:1997:PRB}
E.~Schachinger, J.~P.~Carbotte, and F.~Marsiglio, 
Phys.~Rev.~B \textbf{56}, 2738 (1997). 

\bibitem{Munzar:1999:PhysicaC}
D.~Munzar, C.~Bernhard, and M.~Cardona, 
Physica C \textbf{312}, 121 (1999). 

\bibitem{Carbotte:1999:Nature} 
J.~P.~Carbotte, E.~Schachinger, and D.~N.~Basov, 
Nature (London) \textbf{401}, 354 (1999). 

\bibitem{Abanov:2001:PRB}
A.~Abanov, A.V.~Chubukov, J.~Schmalian, 
Phys. Rev. B \textbf{63}, 180510R (2001).

\bibitem{Casek:2005:PRB}
P.~C\'{a}sek, C.~Bernhard, J.~Huml\'{\i}\v{c}ek, and D.~Munzar, 
Phys. Rev. B \textbf{72}, 134526 (2005).

\bibitem{Hwang:2007:PRB}
J.~Hwang, T.~Timusk, E.~Schachinger, and J.~P.~Carbotte
Phys.~Rev.~B 75, 144508 (2007). 

\bibitem{Nozieres:1999:book}
P.~Nozieres and D.~Pines, \textit{The Theory of Quantum Liquids}
(Perseus Books, Cambridge, Massachusetts, 1999). 

\bibitem{Millis:2003:PRB}
A.J.~Millis, H.D.~Drew, Phys. Rev. B \textbf{67}, 214517 (2003).

\bibitem{Millis:2005:PRB}
A.J.~Millis, A.~Zimmers, R.P.S.M.~Lobo, N.~Bontemps, and C.C.~Homes
Phys. Rev. B \textbf{72}, 224517 (2005).

\bibitem{Monthoux:1994:PRB}
P.~Monthoux, D.~Pines, Phys. Rev. B \textbf{49}, 4261 (1994).

\bibitem{Manske}
D.~Manske, \textit{Theory of Unconventional Superconductors} (Springer-Verlag,
Berlin, 2004).

\bibitem{Kontani:2006:JPSJ}
H.~Kontani, J. Phys. Soc. Jpn. \textbf{75}, 013703 (2006).

\bibitem{Kontani:2005:condmat}
H.~Kontani, cond-mat/0511015 (unpublished).

\bibitem{Benfatto:2005:PRB}
L.~Benfatto, S.G.~Sharapov, N.~Andrenacci, and H.~Beck,
Phys. Rev. B \textbf{71}, 104511 (2005).

\bibitem{Aristov:2005:PRB}
D.N.~Aristov and R.~Zeyher, Phys. Rev. B \textbf{72}, 115118 (2005).

\bibitem{Monthoux:1993:PRB} P.~Monthoux and D.~Pines, 
Phys.~Rev.~B \textbf{47}, 6069 (1993). 

\bibitem{Eschrig:2003:PRB} 
M.~Eschrig and M.R.~Norman,
Phys. Rev. B \textbf{67}, 144503 (2003).

\bibitem{Nambu:1960:PR}
Y.~Nambu, Phys. Rev. \textbf{117}, 648 (1960)

\bibitem{Schrieffer:1988:Book}
J.R.~Schrieffer, \textit{Theory of Superconductivity} (Addison-Wesley,
Reading, MA, 1988).

\bibitem{Andersen:1995:JPCS}
The values of $t$, $t'$ are close to the LDA results published in
O.K.~Andersen, A.I.~Liechtenstein, O.~Jepsen, and F.~Paulsen, 
J. Phys. Chem. Solids \textbf{56}, 1573 (1995).

\bibitem{Fong:2000:PRB} 
H.F.~Fong, P.~Bourges, Y.~Sidis, L.P.~Regnault, J.~Bossy, A.~Ivanov,
D.L.~Milius, I.A.~Aksay, and B.~Keimer,
Phys. Rev. B \textbf{61} 14773 (2000).

\bibitem{Eschrig:2002:PRL}
M.~Eschrig, M.R.~Norman, Phys. Rev. Lett. \textbf{89}, 277005 (2002)

\bibitem{Vidberg:1977:JLTP}
H.J.~Vidberg and J.W.~Serene, J. Low Temp. Phys. \textbf{29}, 179 (1977).

\bibitem{Boris:2004:Science}
A.V.~Boris, N.N.~Kovaleva, O.V.~Dolgov, T.~Holden, C.T.~Lin, B.~Keimer,
and C.~Bernhard, Science \textbf{304}, 708 (2004).

\bibitem{Marel:2003:Nature}
D.~van~der~Marel, H.J.A.~Molegraaf, J.~Zaanen, Z.~Nussinov, 
F.~Carbone, A.~Damascelli, H.~Eisaki, M.~Greven, P.H.~Kes, and
M.~Li, Nature (London) \textbf{425}, 271 (2003).

\bibitem{Timusk:2004:Nature} 
J.~Hwang, T.~Timusk, and D.G.~Gu, Nature (London) \textbf{427}, 714 (2004).

\bibitem{Marsiglio:1998:PhysLett} 
F.~Marsiglio, T.~Startseva, and J.~P.~Carbotte,
Physics Letters A \textbf{245}, 172 (1998). 

\bibitem{Schulga:2001:book}
S.~V.~Schulga, in {\it Material Science, Fundamental 
Properties and Future Electronic Applications of high-Tc Superconductors}, 
edited by S.~L.~Drechsler and T.~Mischonov (Kluwer Academic, Dordrecht, 2001),
pp.~323-360. 

\bibitem{Borisenko:2006:PRL}
S.V.~Borisenko, A.A.~Kordyuk, V.~Zabolotnyy, J.~Geck, D.~Inosov, A.~Koitzsch,
J.~Fink, M.~Knupfer, B.~B\"{u}chner, V.~Hinkov, C.T.~Lin, B.~Keimer, T.~Wolf,
S.G.~Chiuzb\u{a}ian, L.~Patthey, and R.~Follath,
Phys. Rev. Lett. \textbf{96} 117004 (2006).

\bibitem{Schachinger:2005:PRB}
E.~Schachinger and J.P.~Carbotte, Phys. Rev. B \textbf{72}, 014535 (2005).

\end{thebibliography}
\end{document}